\documentclass[sigconf]{acmart}
\usepackage{booktabs} 
\pdfoutput=1
\usepackage{graphicx}
\usepackage{tabularx}
\usepackage{amsmath}
\usepackage[ruled]{algorithm2e}
\usepackage{multirow}
\usepackage{adjustbox}
\graphicspath{{figs/} }
\usepackage{balance}
\usepackage{flushend}
\usepackage{subcaption}
\usepackage{array}
\newcolumntype{P}[1]{>{\centering\arraybackslash}p{#1}}
\newcolumntype{M}[1]{>{\centering\arraybackslash}m{#1}}
\usepackage{comment}
\newtheorem{mydef}{Definition}
\usepackage{caption}
\usepackage{enumitem}
\setcopyright{rightsretained}
\acmDOI{10.475/123_4}
\acmISBN{123-4567-24-567/08/06}
\editor{Jennifer B. Sartor}
\editor{Theo D'Hondt}
\editor{Wolfgang De Meuter}

\begin{document}
\title{MARS: Memory Attention-Aware Recommender System}

\author{Lei Zheng, Chun-Ta Lu}
\affiliation{%
\institution{Department of Computer Science}
  \institution{University of Illinois at Chicago}
  \city{Chicago} 
  \state{IL, U. S. A.} 
  \postcode{60661}
}
\email{lzheng21,clu29@uic.edu}

\author{Lifang He}
\affiliation{%
\institution{Weill Cornell Department of Healthcare Policy \& Research}
  \institution{Cornell University}
  \city{New York} 
  \state{NY, U. S. A.} 
  \postcode{60661}
}
\email{lifanghescut@gmail.com}

\author{Sihong Xie}
\affiliation{%
\institution{Computer Science and Engineering Department}
  \institution{Lehigh University}
  \city{Bethlehem} 
  \state{PA, U. S. A.} 
  \postcode{60661}
}
\email{sxie@cse.lehigh.edu}

\author{Vahid Noroozi, He Huang and Philip S. Yu}
\affiliation{%
\institution{Department of Computer Science}
  \institution{University of Illinois at Chicago}
  \city{Chicago} 
  \state{IL, U. S. A.} 
  \postcode{60661}
}
\email{vnoroo2,hehuang,psyu@uic.edu}

\renewcommand{\shortauthors}{L. Zheng et al.}

\begin{abstract}
In this paper, we study the problem of modeling users' diverse interests. Previous methods usually learn a fixed user representation, which has a limited ability to represent distinct interests of a user. In order to model users' various interests, we propose a Memory Attention-aware Recommender System (MARS). MARS utilizes a memory component and a novel attentional mechanism to learn deep \textit{adaptive user representations}. Trained in an end-to-end fashion, MARS adaptively summarizes users' interests. In the experiments, MARS outperforms seven state-of-the-art methods on three real-world datasets in terms of recall and mean average precision. We also demonstrate that MARS has a great interpretability to explain its recommendation results, which is important in many recommendation scenarios.
\end{abstract}

%
%


\maketitle

\section{Introduction}
\label{sec:intro}
How can we accurately model users' interests? It is a fundamental question for building recommender systems (RS). To answer this question, we observed one essential characteristic of users' interests: \textit{diversity}. Users are interested in different kinds of items and interests of users are diverse. For example, a user purchased several books while she or he also bought an electrical gadget. Furthermore, owing to the diversity of a user's interests, only a subset of the user's purchased products can reveal if the user is interested in another product. For instance, a user may purchase an iPad case because the user bought an iPad rather than a book or a pair of shoes in her last week's shopping list. Hence, it is a nontrivial task to model the diversities of users' interests.

\begin{figure}[t]
\centering\includegraphics[height=0.24\textheight,width=0.44\textwidth]{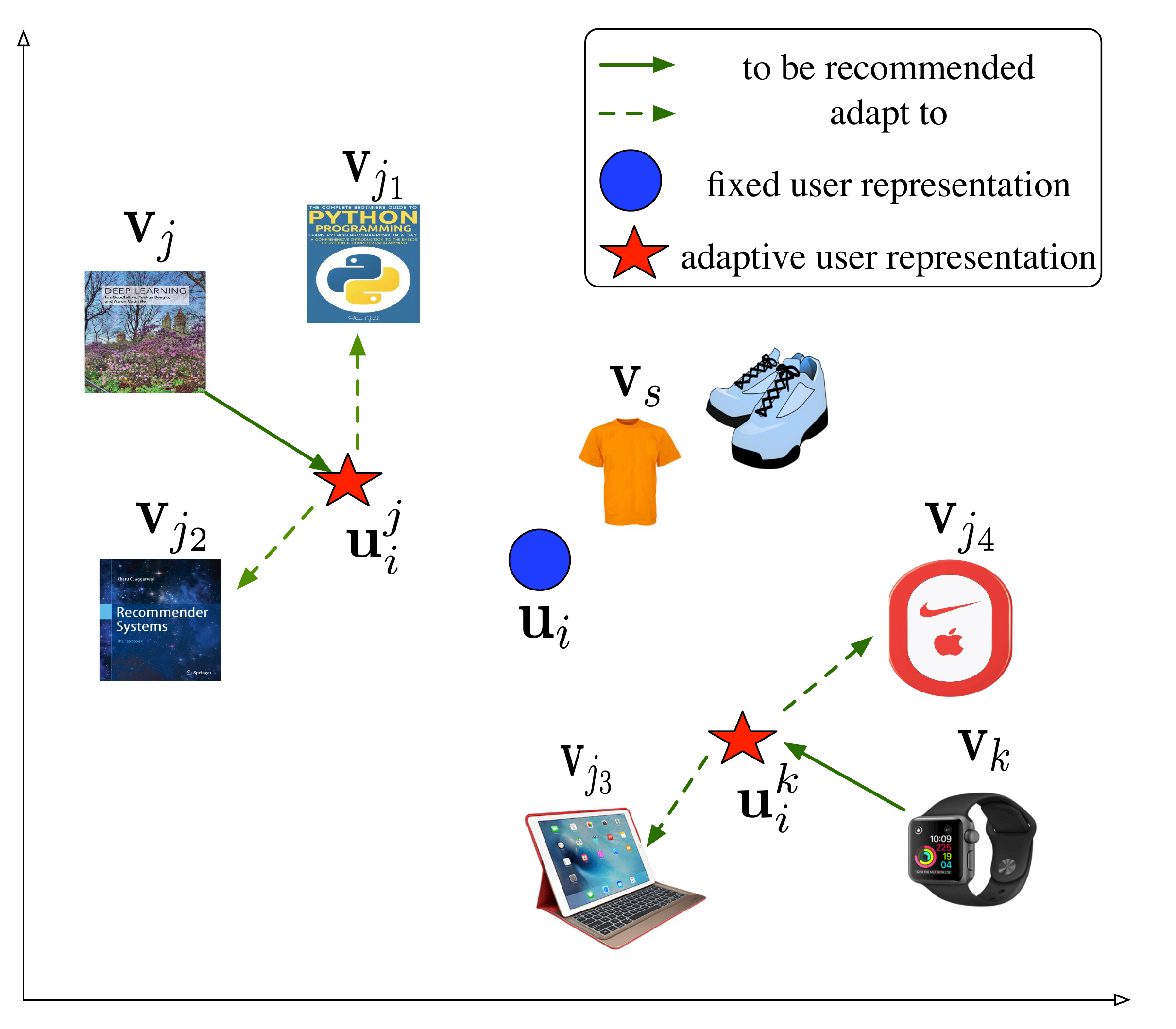}
\caption{A fixed user representation $\mathbf{u}_i$ fails to express user $i$'s diverse interests. \textit{Adaptive user representations} dynamically adapt to relevant items. }
\label{fig::user_embedding}
\centering
\end{figure}

However, how to devise representation vectors to express users' diverse interests is challenging. Existing methods often project users and items into fixed low-dimensional representation vectors in a user-item joint space. We argue that a fixed user representation largely restrains models from accurately modeling users' diverse interests. In the space, similar items are close to each other and the distance between a user and an item implies how much the user is interested in the item. As shown in Figure \ref{fig::user_embedding}, a user $i$ liked a list of different kinds of items $\{j_1,j_2,j_3,j_4\}$. Because of the diversity of items liked by user $i$, their representation vectors ($\mathbf{v}_{j_1}$,$\mathbf{v}_{j_2}$,$\mathbf{v}_{j_3}$ and $\mathbf{v}_{j_4}$) form two clusters in the space. As a result, the fixed user representation $\mathbf{u}_i$ resides between the two clusters in the space. In this case, a system employing fixed user representations will recommend item $s$ to user $i$ instead of item $j$ or item $k$. Although $\mathbf{v}_s$ is closer to $\mathbf{u}_i$ than $\mathbf{v}_j$ or $\mathbf{v}_k$ in the space, it is obvious that either item $j$ or item $k$ is a much more reliable recommendation than item $s$. One may increase the representation dimensionality to overcome the restriction but will make item representations more scattered in the space and cause a huge increase of model parameters.

Moreover, for most of current deep RS, interpreting its recommendations is also difficult and demanding \cite{jannach2010recommender,schafer1999recommender}. 
Despite the effectiveness of representation vectors for predicting interactions, these vectors reside in latent spaces and are not understandable. In order to enhance transparency and trust in the recommendation process, an increasing need of RS is to provide not only accurate but also interpretable recommendations. This is particularly important in a business-to-business setting, where recommendations are generated for experienced sales staff and not directly for the end-client. 

In this paper, in order to model representations of users to express their diverse interests and build a recommendation model capable of interpreting its recommendations, we develop MARS: Memory Attention-Aware Recommender System. First of all, MARS exploits the power of deep learning to learn representations of items from the item content. More importantly, motivated by the observation of user behaviors, we facilitate a memory component and a novel item-level attention mechanism to devise a deep \textit{adaptive user representation}. Unlike fixed user representations, \textit{adaptive user representations} dynamically adapt to locally activated items. As shown in Figure \ref{fig::user_embedding}, for a candidate item $j$, an \textit{adaptive user representation} $\mathbf{u}_i^{j}$ dynamically adapts to locally activated items: $j_1$ and $j_2$. To recommend item $k$, another \textit{adaptive user representation} $\mathbf{u}_i^{k}$ adapts to relevant items like $j_3$ and $j_4$. The concept of \textit{adaptive user representation} can be defined as :
\begin{mydef}
\label{def}
(\textbf{Adaptive User Representation}). For a user $i$ and an item list: $\mathcal{L}=\{j_1,j_2,...,j_{n_i}\}$ with $n_i$ items liked by the user, in order to recommend a candidate item $j$, an adaptive user representation dynamically adapts to items in $\mathcal{L}$ which are highly relevant to item $j$. 
\end{mydef}
The contributions of this work can be summarized as follows:
 \begin{itemize}
 \item \textbf{Adaptive User Representations}: To the best of our knowledge, we are the first to introduce a deep end-to-end recommendation model to learn \textit{adaptive user representations}. Especially, a memory component is utilized to capture users' interests in an end-to-end fashion. An attentional mechanism is facilitated to handle the diversities of users' interests.
\item \textbf{An Interpretable Model}: Benefiting from the item-level attention mechanism, MARS has a good interpretability to explain why an item gets recommended to a user by showing relevant items liked by the user. 
\item \textbf{Strong Performance}: In the experiments, we demonstrate that MARS can significantly outperform strong baselines on three real-world datasets.

   \end{itemize}
  
\section{Background and Preliminaries}
\label{sec:prelim}
In this section, we present the background and preliminaries of this study. Throughout the paper, we denote scalars by either lowercase or uppercase letters, vectors by boldfaced lowercase letters, and matrices by boldfaced uppercase letters.

We consider the most common scenario with implicit feedbacks. For implicit feedbacks, only positive observations, such as clicks, purchase or likes, are available. The non-observed user-item pairs, e.g. a user has not bought an item yet, are a mixture of real negative feedback (the user is not interested in buying the item) and missing values (the user might want to buy the item in the future).

Although, in different recommendation scenarios, the item content can be in distinct formats, such as text or images, we limit our study to recommend items associated with the textual content, which usually describe important characteristics of items. However, note that the proposed model can be easily generalized to recommend different types of items, such as songs or videos.

Let us denote a set of users as $\mathcal{U}$ and an item set $\mathcal{I}$. Each item $j \in \mathcal{I}$ is associated with a document $\boldsymbol{Y}^j = \{w_1,w_2,...,w_{n_j}\}$ containing $n_j$ words. For a user $i \in \mathcal{U}$, let $\mathcal{I}^+_i = \{I_1,I_2,...,I_{n_i}\}$ denote the set of $n_i$ items liked by user $i$ and $\mathcal{I}^-_i$ denote the remaining items. Furthermore, all remaining items liked by user $i$ except item $j$, denoted as $\mathcal{I}^+_i/j$, are utilized to model user representations.

Finally, we define the recommendation problem which we study in this paper as follows:
\begin{mydef}
(\textbf{Problem Definition}). Given user and item sets: $\mathcal{U}$ and $\mathcal{I}$, for each user $i \in \mathcal{U}$, we aim to recommend a ranked list of items from $\mathcal{I}_i^-$ that are of interests to the user.
\end{mydef}

\section{Proposed Model}
\label{sec::model}
The overall architecture of MARS is shown as in Figure \ref{fig::model}. Inspired by the recent success of Convolutional Neural Network (CNN) \cite{lecun1990handwritten} in various tasks \cite{kim2016convolutional,krizhevsky2012imagenet}, we first facilitate two CNN models to serve for different purposes. One CNN model, denoted as $f_{user}(:;\mathbf{\Psi})$ and parameterized by $\mathbf{\Psi}$, is designed to learn a memory component $\mathbf{C}$ from documents associated with items liked by user $i$. The other CNN model, denoted as $f_{item}(:;\mathbf{\Omega})$ and parameterized by $\mathbf{\Omega}$, is responsible for learning an representation $\mathbf{v}_j$ of item $j$. Later, a novel item-level attention mechanism is proposed to build on the top of the memory component to learn a deep \textit{adaptive user representation} $\mathbf{u}^j_i$. At last, with $\mathbf{u}^j_i$ and $\mathbf{v}_j$, user $i$'s preference score of item $j$ can be derived. 
\subsubsection{CNN Models}
\begin{figure}[t]
\centering
\includegraphics[width=0.47\textwidth]{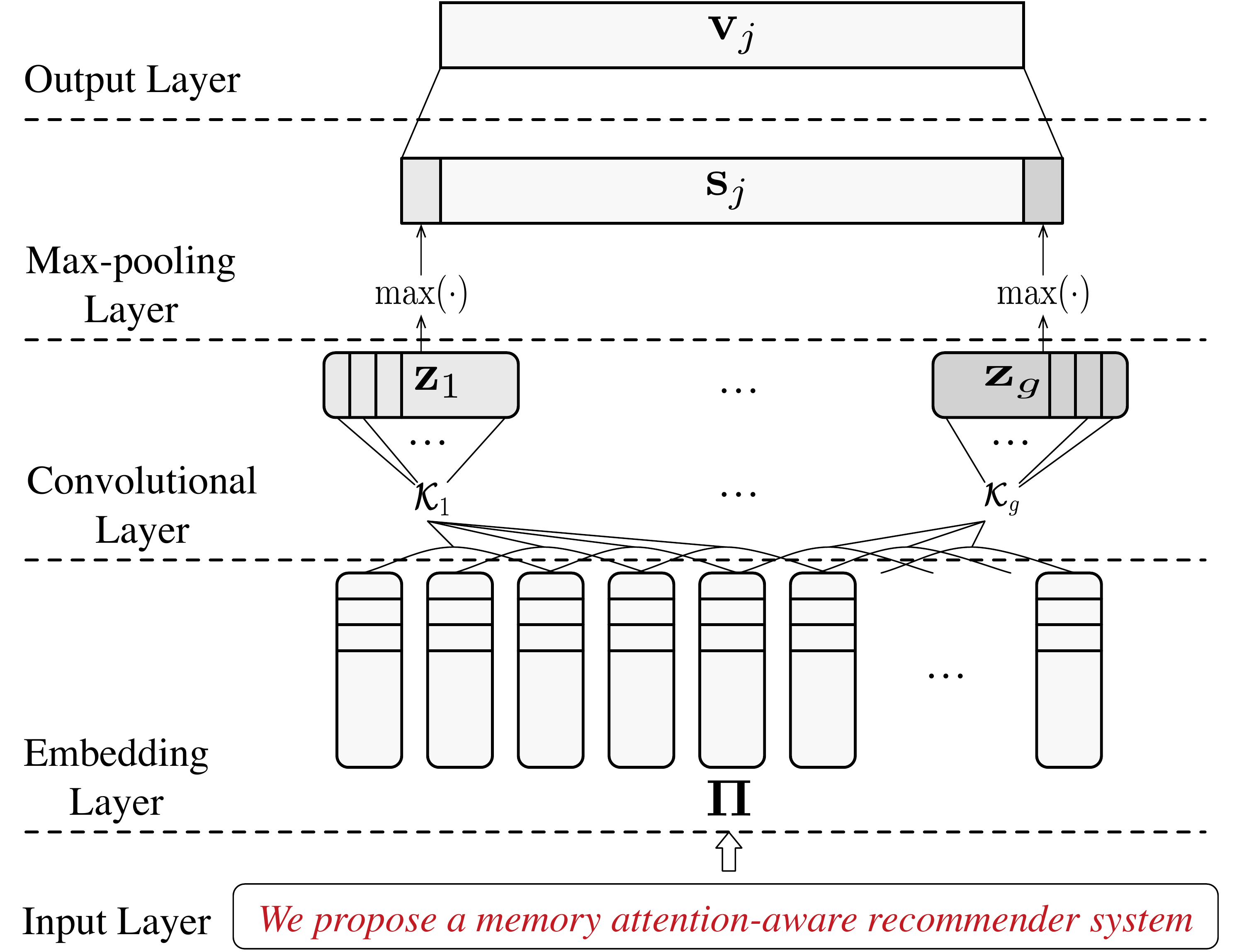}
\caption{Our CNN architecture used for learning item representations.}
\label{fig::cnn}
\centering
\end{figure}

In this subsection, since $f_{user}(:;\mathbf{\Psi})$ and $f_{item}(:;\mathbf{\Omega})$ only differ in their inputs, we focus on illustrating the process of $f_{item}(:;\mathbf{\Omega})$ in detail. The same process is applied for $f_{user}(:;\mathbf{\Psi})$ with similar layers. The architecture of CNN used in this paper is shown in Figure \ref{fig::cnn}.

Given an item $j$ and its associated document $\boldsymbol{Y}^j = \{w_1,w_2,...,w_{n_j}\}$, in order to make use of complex lexical semantics within $\boldsymbol{Y}^j$, we first utilize a word embedding layer to transform the document into a dense numeric matrix $\mathbf{\Pi}$. Formally, the word embedding layer is defined as $\mathcal{H}:\mathcal{V} \rightarrow \mathcal{R}^e$, where $\mathcal{V}$ represents the dictionary of words. We map each word in $\boldsymbol{Y}^j$ into an $e$ dimensional vector as:
\begin{eqnarray}
\mathbf{\Pi}=\mathcal{H}(w_1) \oplus \mathcal{H}(w_2)\oplus,...,\oplus \mathcal{H}(w_{n_j}),
\end{eqnarray}
where $\mathbf{\Pi} \in \mathcal{R}^{e \times n_j}$, each column corresponds to a vector for a word in the document $\boldsymbol{Y}^j$. Note that $\mathbf{\Pi}$ is randomly initialized and is further trained through the optimization process.

Following the word embedding layer, a convolutional layer is built to extract important contextual features that can represent items. A convolution layer consists of $g$ neurons in total, each of which applies convolution operator on $\mathbf{\Pi}$ as:
\begin{equation}
\label{eq2}
\mathbf{z}=ReLU(\mathbf{\Pi} \ast \mathcal{K}+b).
\end{equation}
Here symbol $\ast$ is the convolution operator, $\mathcal{K} \in \mathcal{R}^{e \times c}$ and $b \in \mathcal{R}$ are a convolutional filter and a bias term, respectively, and $c$ denotes the window size of the convolutional filter. We use $ReLU$ \cite{nair2010rectified} as our activation function.

After the convolutional layer, $\mathbf{z} \in \mathcal{R}^{(n_j-c+1) \times 1}$ becomes a vector consisting of $n_j-c+1$ contextual features captured by the convolutional kernel $\mathcal{K}$. To extract the most important feature value from $\mathbf{z}$, we apply a max-pooling operation on $\mathbf{z}$ as shown in Figure \ref{fig::cnn}. With $g$ neurons in the convolutional layer, we obtain a column vector $\mathbf{s}_j$ consisting of $g$ important contextual features as:
\begin{equation}
\mathbf{s}_j = \{max(\mathbf{z}_1),max(\mathbf{z}_2),...,max(\mathbf{z}_g)\}.
\end{equation}
At last, contextual features $\mathbf{s}_j$ of item $j$ is projected to a $K$-dimensional space through a fully connected layer as:
\begin{equation}
\mathbf{v}_j = tanh(\mathbf{\hat{W}}\mathbf{s}_j + \hat{b}),
\end{equation}
where $\mathbf{\hat{W}} \in \mathcal{R}^{K \times g}$ and $\hat{b} \in \mathcal{R}$.
Since the document $\boldsymbol{Y}^j$ is associated with item $j$ and characterizes item $j$, $\mathbf{v}_j \in \mathcal{R}^{K \times 1}$ can be regard as a deep representation of item $j$. 
\subsection{Memory Component} 
A set of items liked by a user naturally reflects the user's interests. However, it is not easy to summarize diverse items liked by user $i$ into a user representation. As discussed in the introduction section, a fixed user representation fails to convey diverse interests of a user. 

Different from previous methods \cite{wang2016collaborative,cheng2016wide}, in order to learn a deep and accurate user representation, as shown in Figure \ref{fig::model}, we first propose to first utilize $f_{user}(:;\mathbf{\Psi})$ to learn a memory component for items in $\mathcal{I}^+_i/j$ as: 
\begin{eqnarray}
\label{memory}
\textbf{C}=&\{f_{user}(\boldsymbol{Y^1};\mathbf{\Psi}),f_{user}(\boldsymbol{Y^2};\mathbf{\Psi}),...,f_{user}(\boldsymbol{Y}^{n_i-1};\mathbf{\Psi})\} \nonumber \\
=&\{\mathbf{v}_1,\mathbf{v}_2,...,\mathbf{v}_{n_i-1}\},
\end{eqnarray} 
Each column of $\mathbf{C} \in \mathcal{R}^{K \times (n_i-1)}$ corresponds to a representation of an item in $\mathcal{I}_i^+/j$. Hence, the matrix $\mathbf{C}$ characterizes the user's interests.
\subsubsection{Item-level Attention}
\begin{figure}[t]
\centering
\includegraphics[height=0.3\textheight,width=0.45\textwidth]{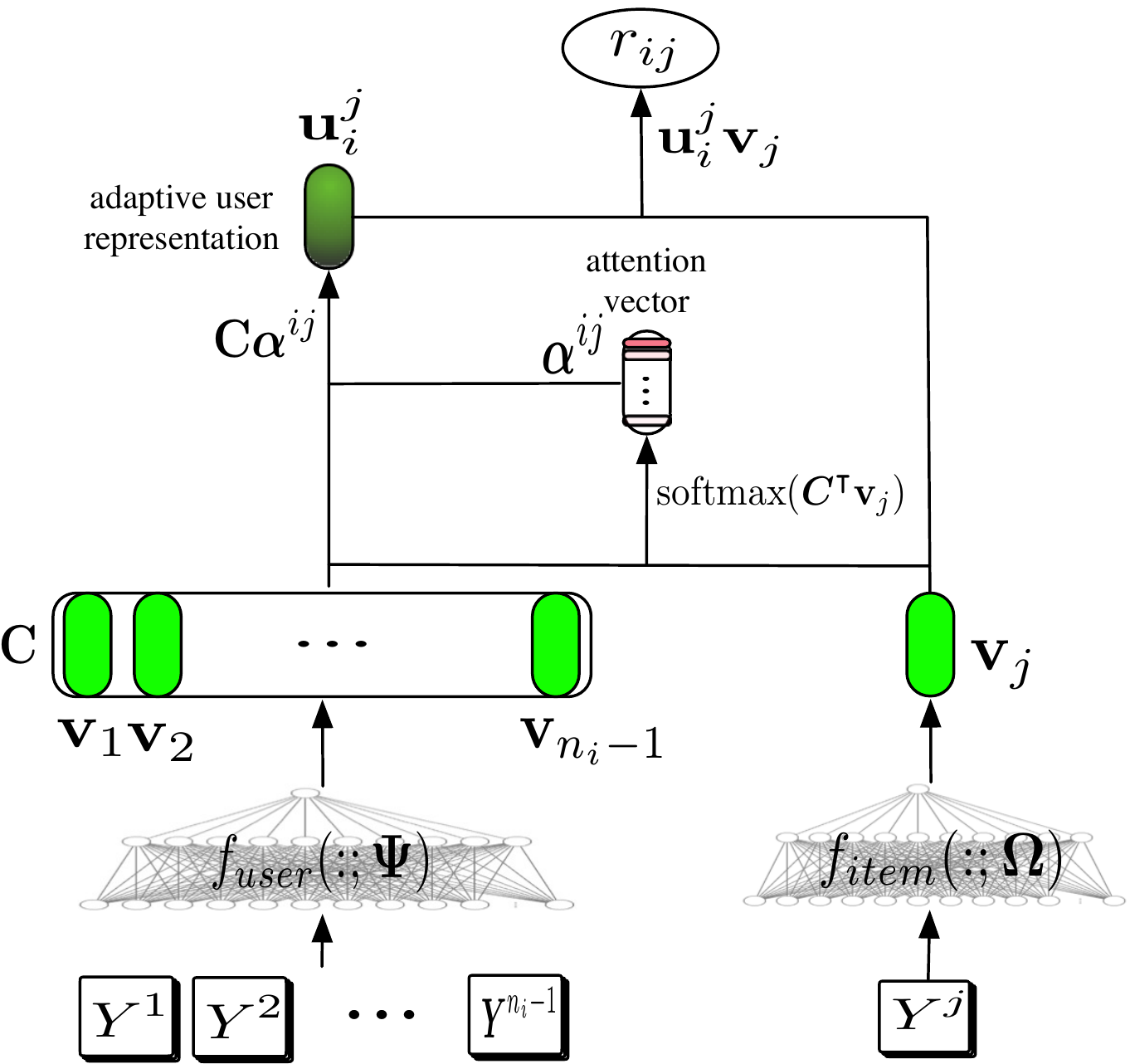}
\caption{The architecture of MARS.}
\label{fig::model}
\centering
\end{figure}

Although the memory component $\mathbf{C}$ stores all item information of user $i$, our goal is to learn a deep representation of user $i$ based on $\mathbf{C}$. As discussed in the Introduction section, items in $\mathcal{I}^+_i/j$ are diverse and only a subset of $\mathcal{I}^+_i/j$ is relevant to item $j$.  
Motivated by the intuition above, we introduce an item-level attention mechanism on the memory component to capture items in $\mathcal{I}^+_i/j$ relevant to item $j$. To do so, as shown in Figure \ref{fig::model}, for a given item $j$, we first feed its corresponding document $\boldsymbol{Y}^j$ into $f_{item}(:;\mathbf{\Omega})$ to derive an item representation as: 
\begin{equation}
\label{q_component}
\textbf{v}_j = f_{item}(\boldsymbol{Y}^j;\mathbf{\Omega}),
\end{equation}
where $\mathbf{v}_j \in \mathcal{R}^{K \times 1}$ is a column vector serving as an representation of item $j$.

More items in $\mathcal{I}^+_i/j$ with high relevance scores or locally activated to item $j$ mean that the user $i$ is more likely interested in $j$. Thus, with the memory component $\mathbf{C}$ and $\mathbf{v}_j$, we measure the relevance score or activation degree between $\mathbf{v}_j$ and each item representation in $\mathbf{C}=\{\textbf{v}_1,\textbf{v}_2,...,\textbf{v}_{n_i-1}\} \in \mathcal{R}^{K \times (n_i-1)}$ by taking the inner product followed by a softmax as:
\begin{equation}
\label{attention}
\boldsymbol{\alpha}^{ij} = softmax({\boldsymbol{C}^\intercal\boldsymbol{v}_j}).
\end{equation}
Defined in this way, $\boldsymbol\alpha^{ij} \in \mathcal{R}^{(n_i-1)\times1}$ is an attention column vector of user $i$ for item $j$. A larger value in $\boldsymbol\alpha^{ij}$ indicates higher relevance between item $j$ and a corresponding item in $\mathcal{I}^+_i/j$. A larger value means a higher weight for deriving interests of user $i$ for item $j$.
\subsubsection{Deep Adaptive User Representations} A fixed user representation is unable to express diverse interests of the user. In order to uncover a user's interests in a candidate item $j$, we introduce a deep \textit{adaptive user representation}. The proposed user representation is dependent on the candidate item $j$ and non-fixed. By focusing on items in $\mathcal{I}^+_i/j$ which are highly activated, it is able to model users' diverse interests.

With the matrix $\mathbf{C}$ and the attention vector $\boldsymbol\alpha^{ij}$, a deep attention-aware \textit{adaptive user representation} is formed by a weighted sum of $\mathbf{C}=\{\mathbf{v}_1,\mathbf{v}_2,...,\mathbf{v}_{n_i-1}\}$ as:
\begin{equation}
\label{user_embedding}
\mathbf{u}^j_i = \mathbf{C}{\boldsymbol\alpha^{ij}}.
\end{equation}
In Eq.~(\ref{user_embedding}), the attention vector $\boldsymbol\alpha^{ij}$ weights each item representation in $\mathbf{C}$. An \textit{adaptive user representation} $\mathbf{u}^j_i$ is obtained by adaptively focusing on items in $\mathcal{I}^+_i/j$ activated by item $j$. 
\subsection{Pair-wise Learning and Prediction}
Recall that an \textit{adaptive user representation} $\mathbf{u}^j_i$ is obtained by placing an attentional vector on our memory component. Given the representation of a candidate item $\mathbf{v}_j$, one can compute the user $i$'s preference score over item $j$ as: 
\begin{equation}
r_{ij} = {\textbf{u}_i^j}^\intercal\textbf{v}_j.
\end{equation}
Here, we derive a preference score $r_{ij}$ that reflects the preference of the user for the item. Usual approaches for item recommendations are to rank items by sorting them according to the scores. However, these approaches treat recommendation tasks as regression problems and are not optimized for ranking. To train MARS optimized for ranking, inspired by BPR \cite{rendle2009bpr}, we formalize the training data $\mathcal{D}$ as: 
\begin{equation}
\mathcal{D} = \{(i,\mathcal{I}^+_i/j,j,j')|i \in \mathcal{U} \wedge j \in \mathcal{I}^+_i \wedge j' \in \mathcal{I}^-_i\},
\end{equation}
where $i$, $j$ and $j'$ are uniformly sampled form $\mathcal{U}$, $\mathcal{I}^+_i$ and $\mathcal{I}^-_i$, respectively. Note that for each quadruple $(i,\mathcal{I}^+_i/j,j,j')$ in $\mathcal{D}$, $\mathcal{I}^+_i/j$ is a different sampled subset of $\mathcal{I}^+_i$. The sampling strategy is a common practice for pair-wise recommendation learning. \cite{rendle2009bpr} also shows that training on all item pairs can result in slow and poor convergence.

Similar to BPR, instead of scoring single items, we use item pairs to model a user $u$'s preference of item $j$ over item $j'$ as:
\begin{eqnarray}
\label{loss1}
\mathcal{L}&=-\frac{1}{|\mathcal{D}|}\sum _{ (i,\mathcal{I}_i^+/j,j,j')\in \mathcal{D} }^{  }{  \{\ln( \sigma ({{ { \mathbf{u} } }^j_{ i }}^\intercal{ { \mathbf{v} } }_{ j }-{{ { \mathbf{u} } }^{j'}_{ i }}^\intercal{ { \mathbf{v} } }_{ j' }) ) } \nonumber \\
&+\lambda_u{\mathbf{u}_i^j}^\intercal{\mathbf{u}_i^j} +
\lambda_u{\mathbf{u}_i^{j'}}^\intercal{\mathbf{u}_i^{j'}} +
\lambda_v{\mathbf{v}_j}^\intercal{\mathbf{v}_j} + \lambda_v{\mathbf{v}_{j'}}^\intercal{\mathbf{v}_{j'}}\}, 
\end{eqnarray}
where $\textbf{u}^j_i$ and $\textbf{u}^{j'}_i$ denote \textit{adaptive user representations} of user $i$ for item $j$ and item $j'$; $\textbf{v}_j$ and $\textbf{v}_{j'}$ stand for representations of item $j$ and item $j'$; $\lambda_u$ and $\lambda_v$ are regularization terms. The sigmoid function $\sigma$ maps user  $i$'s preference score of item $j$ over item $j'$ into probabilities. 

MARS is trained by minimizing Eq.~(\ref{loss1}). 
The derivatives of parameters in different layers can be computed by applying differentiation chain rule \cite{rumelhart1988learning}. We optimize the model through RMSprop \cite{tieleman2012lecture} over a batch of tuple $\{i,\mathcal{I}^+_i/j,j,j'\}$. 
\begin{algorithm}[t]
 \caption{Training procedure of MARS}
 \label{alg}
\SetAlgoLined {\small
\KwIn{Training set: $\mathcal{D} := \{(i,\mathcal{I}^+_i/j,j,j')|i \in \mathcal{U} \wedge j \in \mathcal{I}^+_i \wedge j' \in \mathcal{I}^-_i\}$,
number of epochs $T$,
batch size $m$, window size $c$, number of convolutional neurons $g$}
\KwOut{Model's parameter set: $\mathbf{\Theta}=\{\mathbf{u},\mathbf{v},\mathbf{\Omega}\}$}
\For{${t}=1,2,\cdots,T$}{
Generate the $t^\textrm{th}$ batch of size $m$ by uniformly sampling from $\mathcal{U}$, $\mathcal{I}_i^+$ and $\mathcal{I}_i^-$;\\
Calculate the memory component $\mathbf{C}$ according to Eq. (\ref{memory})\;
Calculate $\mathbf{v}_j$ and $\mathbf{v}_{j'}$ according to Eq. (\ref{q_component})\;
Calculate $\boldsymbol\alpha^{ij}$ and $\boldsymbol\alpha^{ij'}$ according to Eq.~(\ref{attention})\;
Calculate $\mathbf{u}_i^j$ and $\mathbf{u}_i^{j'}$ according to Eq.~(\ref{user_embedding})\;
Calculate $\mathcal{L}$ according to Eq.~(\ref{loss1})\;
Estimate gradients $\frac{{\partial \mathcal{L}^{}}}{{\partial \mathbf{\Theta}_t }}$ by back propagation\;
Calculate $\mathbf{\Theta}_{t+1}$ with RMSprop \cite{tieleman2012lecture}\;
}
\KwRet{$\mathbf{\Theta}_{T}$};}
\end{algorithm}

Overall, the training procedure of MARS is summarized in Algorithm \ref{alg}. During the test, given a user $i$'s past likes $\mathcal{I}^+_i$, the final item recommendation list for user $i$ is given according to the following ranking criterion:
\begin{equation}
i: j_1 \succcurlyeq j_2 \succcurlyeq ... \succcurlyeq j_n \Rightarrow r_{i,j_1} > r_{i,j_2} > ... > r_{i,j_n}.
\end{equation}
\section{Experiments}
\label{sec::exp}
\subsection{Datasets}
We evaluate the proposed method on three real-world datasets with different kinds of items as the following: 
\begin{itemize}
\item \textit{\textbf{Yahoo! Movies}}: it consists of users rating movies on a scale of 1-5 with a short synopsis. To be consistent with the implicit feedback setting, as in \cite{he2017neural}, we extract only positive ratings (rating 5) for training and testing. After removing movies without a synopsis and users with less than 3 ratings, we obtain a dataset that contains 7,642 users, 11,915 items, and 221,367 positive ratings with 0.24\% density.
\item \textbf{\textit{Amazon Video Games}}: it is collected by \cite{mcauley2015image,he2016ups} and contains game descriptions and ratings from 1 to 5. Similarly, we transformed it into implicit data, where each entry is marked as 0 or 1 indicating whether the user has rated the games. After removed games with descriptions and users with less than 10 ratings, the resulting dataset contains 47,063 video games and 2,670 users with 0.037\% density.
\item \textbf{\textit{Amazon Movies and TV}}: this dataset is also collected by \cite{mcauley2015image,he2016ups}. Similarly, items in this dataset are movies and TV shows available on \textit{Amazon.com}. Users rate them from 1 to 5. In order to transform explicit ratings into implicit feedbacks, we only preserve ratings with the value of 5 and treat them as positive feedbacks. Other entries are marked as negative ones. By removing users with less than 10 ratings, we obtain a dataset with 22,147 users, 178,086 items, and 0.0128\% density. Note that, compared with the previous two datasets, this is a much harder dataset due to its sparsity.
\end{itemize}

The synopses and descriptions of items serve as item contents. After removing stop words and frequencies of the word less than 5, the vocabulary sizes of \textit{Yahoo! Movies}, \textit{Amazon Video Games} and \textit{Amazon Movies and TV} are 33,195, 25,035 and 68,919, respectively. All three datasets are publicly available \footnote{\textit{Yahoo! Movies} can be downloaded at https://webscope.sandbox.yahoo.com; \textit{Amazon Video Games} and \textit{Amazon Movies and TV} are available at https://snap.stanford.edu/data/web-Amazon.html}.
\subsection{Baselines}
\label{baselines}
To validate the effectiveness of MARS, we compare MARS with seven state-of-the-art baseline models. Among them, BPR and NCF  completely ignore the textual content associated with items and all other baselines utilize the content information to boost their performances.
\begin{itemize}
\item \textbf{BPR} \cite{rendle2009bpr}\footnote{Code: \url{https://github.com/zenogantner/MyMediaLite}}:  We use \textbf{B}ayesian \textbf{P}ersonalized \textbf{R}anking based Matrix Factorization, which is based on users' pair-wise preference, as the single collaborative filtering method. BPR completely ignores the usage of item content.

\item \textbf{NCF} \cite{he2017neural}\footnote{Code: \url{https://github.com/hexiangnan/neural_collaborative_filtering}}: \textbf{N}eural \textbf{C}ollaborative \textbf{F}iltering fuses matrix factorization and Multi-Layer Perceptron (MLP) to learn from user-item interactions. The MLP endows
NCF with the ability of modelling non-linearities.

\item \textbf{CMF} \cite{singh2008relational}\footnote{Code: \url{https://github.com/david-cortes/cmfrec}}: \textbf{C}ollective \textbf{M}atrix \textbf{F}actorization simultaneously
factorizes multiple matrices to incorporate different sources of information. We factorize the rating matrix and a matrix consisting of \textit{bag-of-words} features. Note that we follow the work of \cite{hu2008collaborative} to adapt CMF for implicit feedbacks.

\item \textbf{CTR} \cite{wang2011collaborative}\footnote{Code: \url{https://github.com/blei-lab/ctr}}: \textbf{C}ollaborative \textbf{T}opic \textbf{R}egression is based on topic modeling techniques and shows very good performance on recommending articles. 

\item \textbf{DeepFM} \cite{gup2017deepfm}\footnote{Code: \url{https://github.com/ChenglongChen/tensorflow-DeepFM}}: This baseline combines the power of factorization machine \cite{rendle:tist2012}
for recommendations and deep learning for feature learning in a neural network. We concatenate a user id and \textit{bag-of-words} features of each item for training and predictions. 

\item \textbf{CDL} \cite{wang2015collaborative}\footnote{Code: \url{http://www.wanghao.in/code/cdl-release.rar}}: \textbf{C}ollaborative \textbf{D}eep \textbf{L}earning is a recently proposed deep recommender system. It tightly couples a Bayesian formulation of the stacked denoising auto-encoders and Probabilistic Matrix Factorization (PMF) \cite{salakhutdinov2011probabilistic}. The middle layer of auto-encoders serves as a bridge between auto-encoders and probabilistic matrix factorization.

\item \textbf{Wide \& Deep} \cite{cheng2016wide}\footnote{Code: \url{https://www.tensorflow.org/tutorials/wide_and_deep}}: This model is proposed to jointly train wide linear models and deep neural networks to combine the benefits of memorization
and generalization for recommender systems. Similar to DeepFM, a user id and \textit{bag-of-words} features of each item are combined to be fed into both the "wide" and "deep" parts of the model.
\end{itemize}
All baseline models can be categorized into three groups: (1) \textbf{BPR} and \textbf{NCF}: these two models learn only from users' implicit feedback and ignore the textual content; (2) \textbf{CMF} and \textbf{CTR}: this group includes two "shallow" recommendation models; (3) \textbf{DeepFM}, \textbf{CDL} and \textbf{Wide \& Deep}: three state-of-the-art deep learning based models are included to be compared with MARS.
\subsection{Experimental Setup}
\label{sec::exp_setting}
Following \cite{wang2016collaborative}, we evaluate the models on held-out user-item likes. For each dataset, to evaluate MARS and baselines in a sparse setting, we randomly select only 30\% items associated with each user to form the training set. All the remaining are split evenly to serve as the validation and test set. We repeat the evaluation five times with different randomly selected training sets. The average performances are reported in the following sections. For each dataset, to achieve the best performance for each individual model, we conducted carefully parameter study in Section \ref{parameter_study}.

Since we expect RS to not only be able to retrieve relevant items out of all available items but also provide a ranking where items of users' interests are ranked in the top. Therefore, for evaluation, we use two metrics to evaluate the proposed model and the baselines. The first metric we use is recall@$N$. The recall is often used to measure how well a model can retrieve relevant items out of all available items. The recall@$N$ for each user is then defined as:
\begin{equation}
\text{recall@}N = \frac{\# \text{ items the user likes among the top N}}{\text{total number of items the user likes}}.
\end{equation}
Another evaluation metric we use is Mean Average Precision (MAP). The MAP is employed to measure the ranking performances of MARS and baselines. The definition of MAP is given as:
\begin{eqnarray}
\text{MAP} = \frac{1}{|\mathcal{U}|}\sum_{i \in \mathcal{U}}AveP(i), \nonumber \\
AveP(i) = \frac{1}{|K'|} \sum_{k=1}^{K'}P_i(k)rel_i(k),
\end{eqnarray}
where $P_i(k)$ represents the precision of the top $k$ products recommended to user $i$; $rel_i(k)$ denotes whether the $k_{th}$ item has interacted with user $i$ in the test set. Similar to \cite{van2013deep}, we set the cut-off point $K'$ as 500 for each user. 
\subsection{Hyper-Parameter Study}
\label{parameter_study}
Although different models have different hyper-parameters, some hyper-parameters are common and play important roles on model performances. In this section, we optimize the performances of MARS and the baselines by studying the impacts of several important hyper-parameters on the validation sets of \textit{Yahoo! Movies} and \textit{Amazon Video Games}. The learning rate and batch size are empirically set as $0.001$ and $512$ for MARS. All the other hyper-parameters for baselines follow the original papers.
\subsubsection{Dimension of Latent Vectors}
\begin{figure}[t]
\centering
\begin{center}
\begin{subfigure}{0.23\textwidth}
\includegraphics[width= \textwidth,height=.14\textheight]{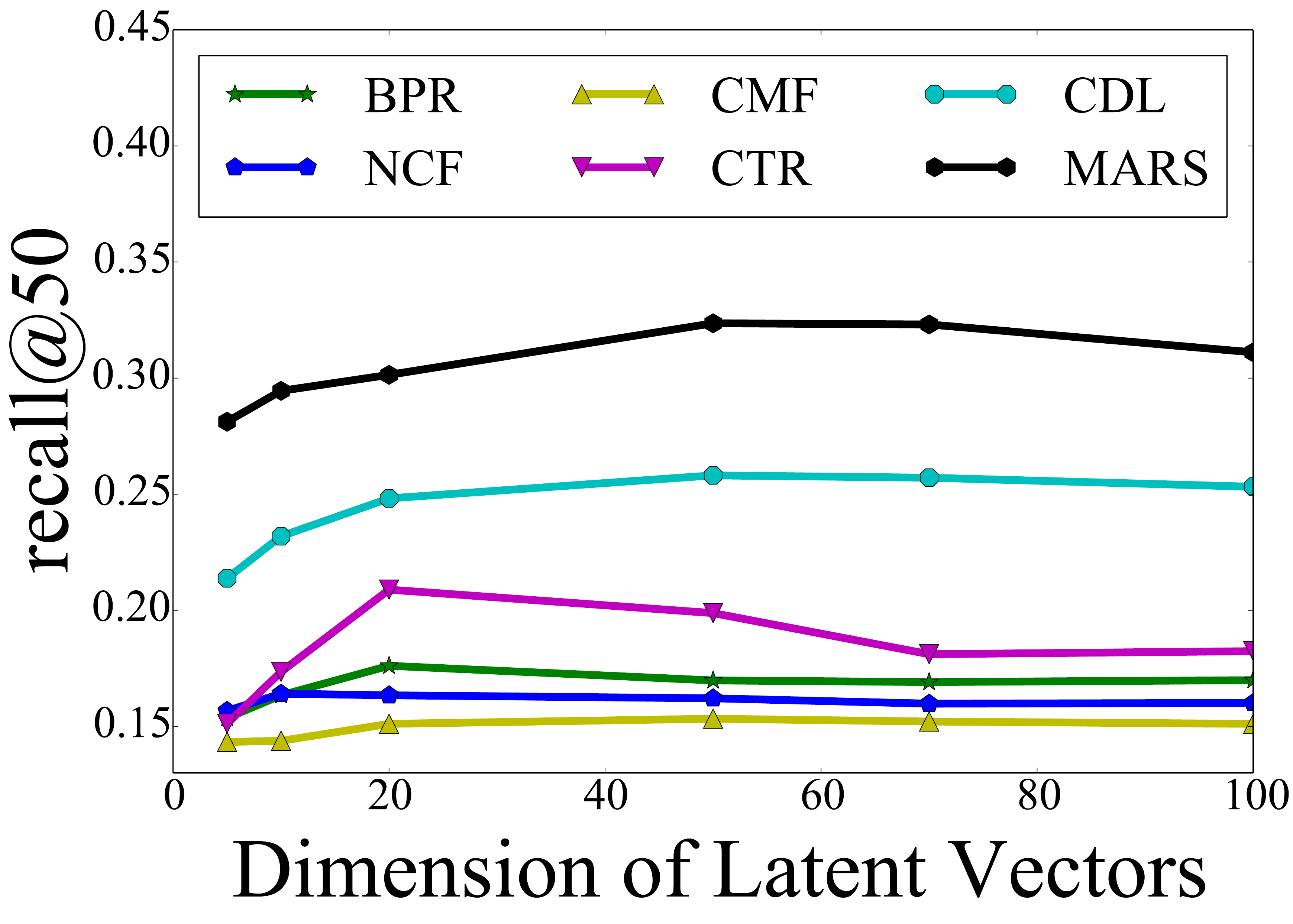}
\caption{Yahoo! Movies}
\end{subfigure}
\begin{subfigure}{0.23\textwidth}
\centering
\includegraphics[width= \textwidth,height=.14\textheight]{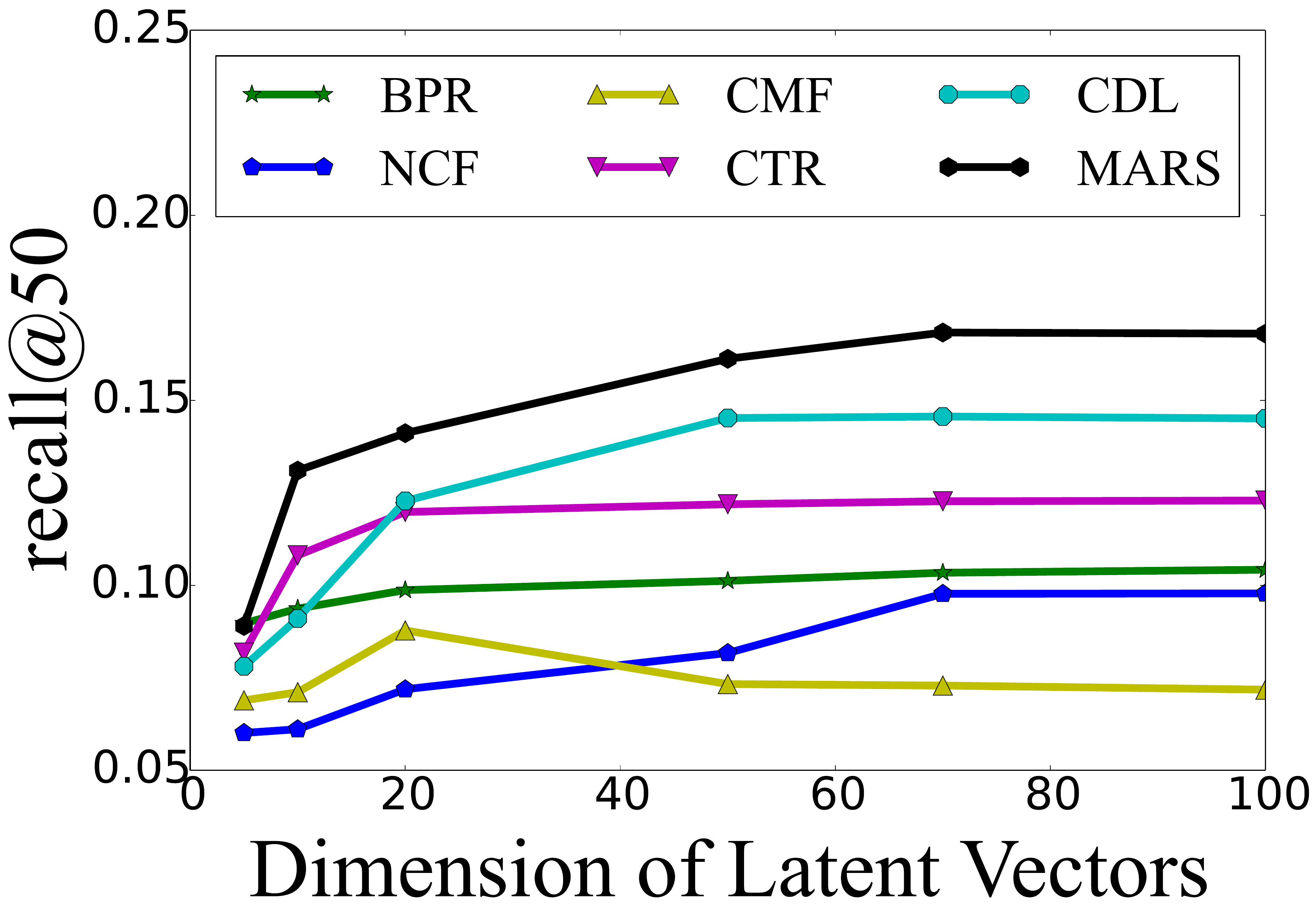}
\caption{Amazon Video Games}
\end{subfigure}
\end{center}
\vspace{-1em}
\caption{The dimension of latent vectors of users and items is varied from 5 to 100 to investigate its impacts on the performances.}
\vspace{-1em}
\label{dimen}
\end{figure}

A low-dimensional latent vector of users and items has a limitation of modeling complex user-item interactions. However, a high-dimensional vector may harm the generalization of the model and increases the number of parameters. In order to optimize the performances of MARS as well as the baselines, we conduct an experiment to investigate the impacts of the dimension of latent vectors. The dimension of latent vectors is searched from $[5,10,20,50,70,100]$ via validation sets from \textit{Yahoo! Movies} and \textit{Amazon Video Games}. As shown in Figure \ref{dimen}, different models reach their own best performances at different dimensions of latent vectors. MARS achieves its best performances when its dimensions are set to $50$ and $70$ on the dataset of \textit{Yahoo! Movies} and \textit{Amazon Video Games}, respectively. 
\subsubsection{Regularization Terms}
\begin{figure}[t]
\centering
\begin{center}
\begin{subfigure}{0.23\textwidth}
\includegraphics[width= \textwidth,height=.14\textheight]{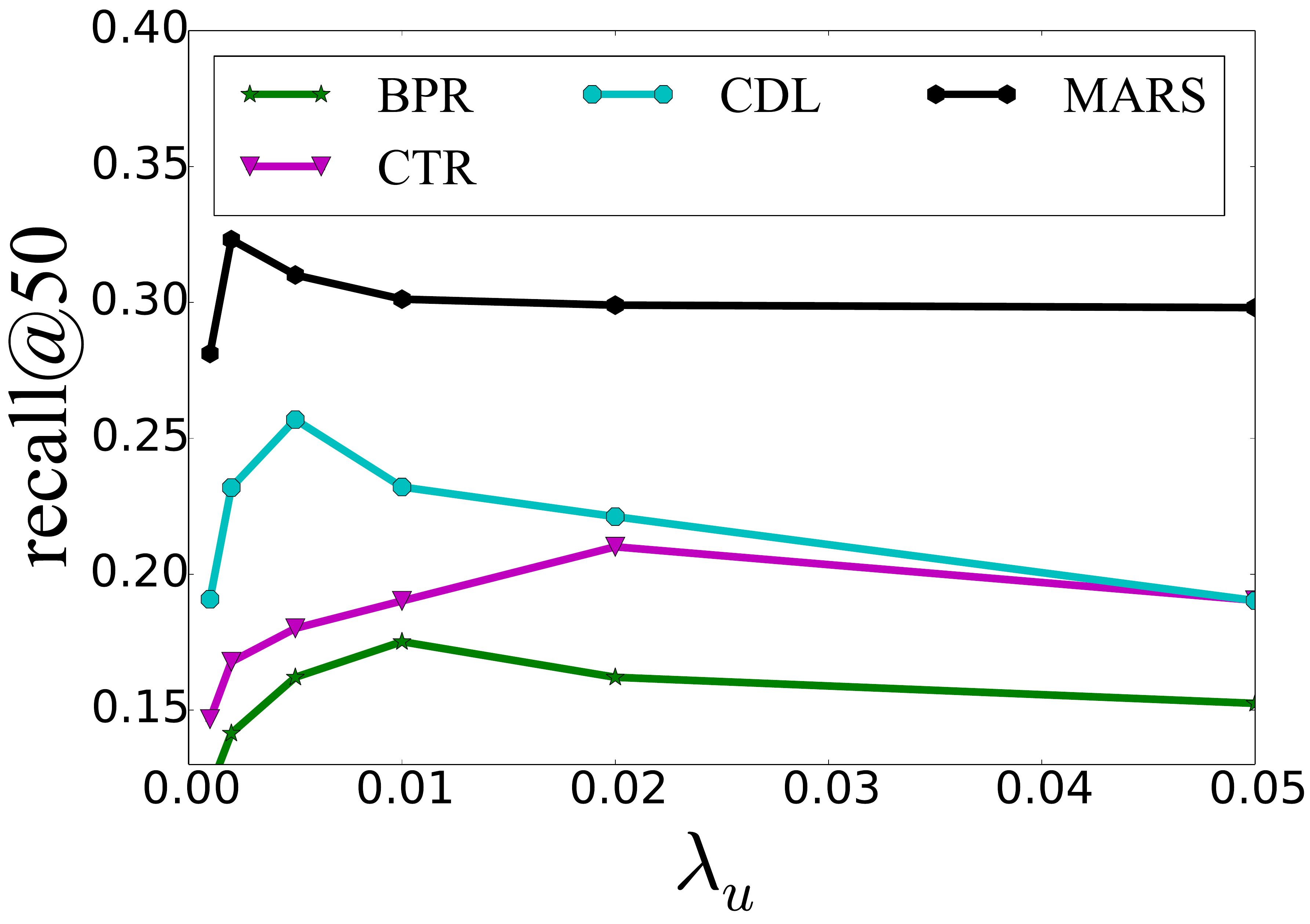}
\caption{Yahoo! Movies}
\end{subfigure}
\begin{subfigure}{0.23\textwidth}
\centering
\includegraphics[width= \textwidth,height=.14\textheight]{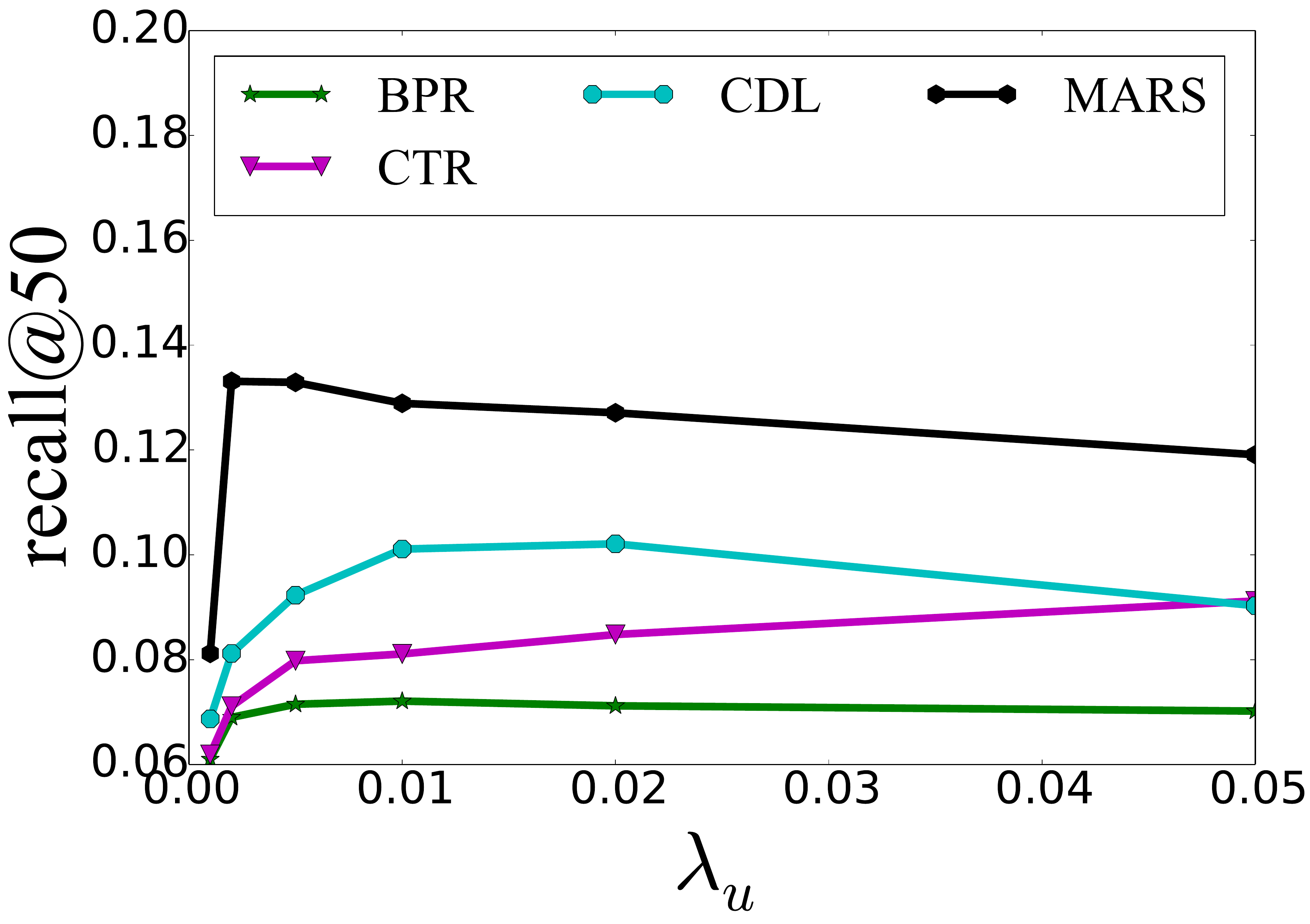}
\caption{Amazon Video Games}
\end{subfigure}
\end{center}
\vspace{-1em}
\caption{The $\mathcal{L}_2$ regularization term $\lambda_u$ is varied from 0.001 to 0.05 to investigate its impacts on the performances.}
\vspace{-1em}
\label{lambda_u}
\end{figure}

\begin{figure}[t]
\centering
\begin{center}
\begin{subfigure}{0.23\textwidth}
\includegraphics[width= \textwidth,height=.14\textheight]{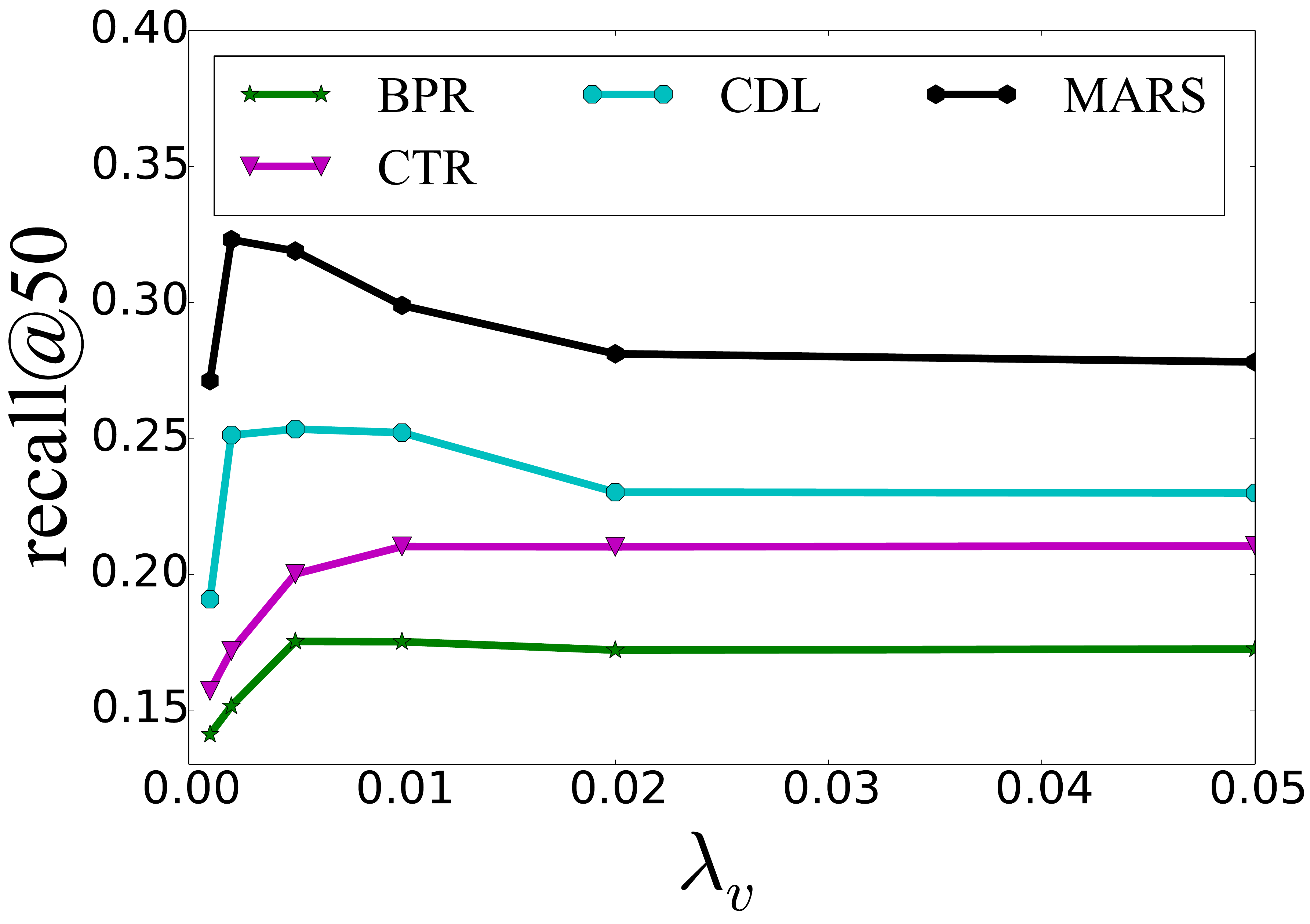}
\caption{Yahoo! Movies}
\end{subfigure}
\begin{subfigure}{0.23\textwidth}
\centering
\includegraphics[width= \textwidth,height=.14\textheight]{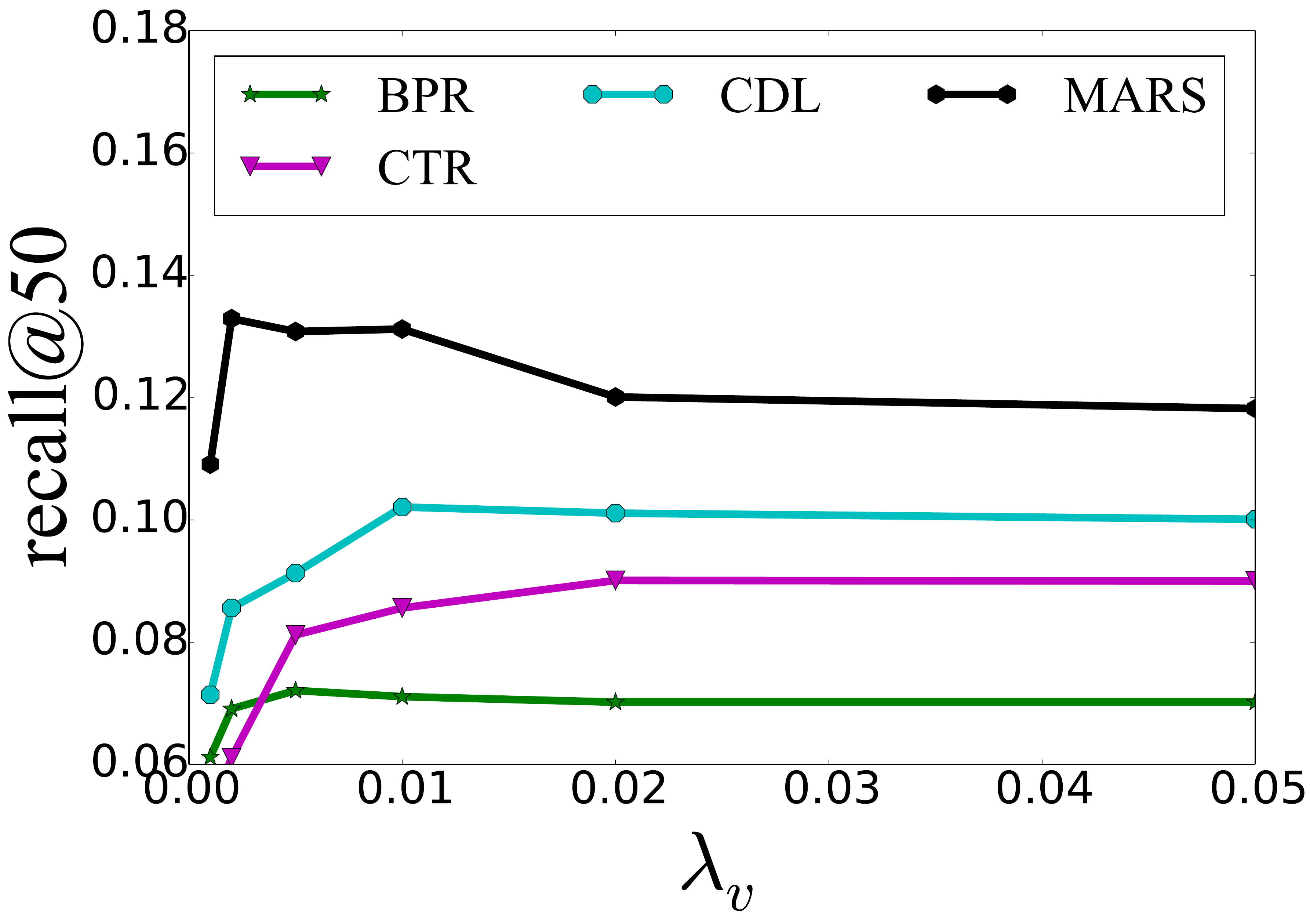}
\caption{Amazon Video Games}
\end{subfigure}
\end{center}
\vspace{-1em}
\caption{The $\mathcal{L}_2$ regularization term $\lambda_v$ is varied from 0.001 to 0.05 to examine its impacts on the performances.}
\vspace{-1em}
\label{lambda_v}
\end{figure}

In order to combat the over-fitting problem, many models place $\mathcal{L}_2$ regularization terms ($\lambda_u$ and $\lambda_v$) on the representation vectors of users and items. We search the two hyper-parameters from $[0.001,0.002,0.005,0.01,0.02,0.05]$ to optimize performances of MARS and the baselines. In Figure \ref{lambda_u} and \ref{lambda_v}, we can see that MARS reaches its best performances when both $\lambda_u$ and $\lambda_v$ are set to $0.002$ on the two datasets. Note that instead of $\mathcal{L}_2$ regularization, NCF and DeepFM employ Dropout \cite{srivastava2014dropout} to prevent over-fitting. And, Wide \& Deep uses the $\mathcal{L}_1$ regularization.  

\subsubsection{Network Architecture}
To find a good network shape for MARS, we also investigate two hyper-parameters: $e$ and $g$, each of which denotes the dimension of word embeddings and the number of convolutional filters, respectively. We perform a grid search on the two hyper-parameters on the validation set of \textit{Yahoo! Movies}. We found MARS can achieve good performances when we set $e=300$ and $g=64$. 
For NCF, as suggested in the original paper \cite{he2017neural}, we employ a three-layer MLP with the shape of $(32,16,8)$. In Wide \& Deep, a three-layer MLP with the shape of $(1024,512,256)$ is used in the "deep" part. In DeepFM, as in \cite{gup2017deepfm}, we build a MLP network with the shape of $(200,200,200)$.
\subsubsection{Other Hyper-Parameters}
For CMF, \textit{bag-of-words} vectors as in \cite{wang2015collaborative} for each item forms a matrix to be simultaneously factorized with the rating matrix. By performing the grid search, we optimize the performances of CMF by setting the weights for the rating matrix and content matrix to 5 and 1. $\alpha$ of CTR is set as $0.1$. To optimize the performance of CDL, we perform a grid search on the hyper-parameters: $\lambda_n$, $\lambda_w$ and $L$.
\subsection{Experimental Results}
\label{sec::results}
\begin{table*}[t]
\centering
\caption{Performance comparison with baselines. The best performance is indicated in \textbf{bold} (higher is better). The standard
deviation is shown in parentheses.}
\label{baseline}
\begin{tabular}{|c|c||M{1cm}|M{1cm}|M{1cm}|M{1cm}|M{1cm}|M{1cm}|M{1cm}|M{1cm}|M{1cm}|}
\hline
\multicolumn{2}{|c||}{Dataset}                                                                                         & BPR&NCF & CMF &CTR  & DeepFM & CDL&Wide \& Deep &MARS& \textit{MARS vs. best}\\ \hline
\multirow{2}{*}{\textit{Yahoo! Movies}}                                                    & recall@50           & 0.1756 (0.001)   &0.1649 (0.008)   & 0.1532 (0.002)         &  0.2067 (0.002)  & 0.2416 (0.002) & 0.2534 (0.001)  & 0.2611 (0.001) &    \textbf{0.3230 (0.001)}      & 23.7\% \\ \cline{2-11} 
                                                                                           & \multicolumn{1}{c||}{MAP} &    0.1045 (0.001) & 0.0973 (0.002) &   0.0894 (0.003)       &     0.1223 (0.002) &  0.1389 (0.001) &   0.1452 (0.001)    & 0.1523 (0.002)  &  \textbf{0.1692 (0.002)}    & 11.1\%  \\ 
                                                      \hline\multirow{2}{*}{\textit{\begin{tabular}[c]{@{}c@{}}Amazon \\Video Games\end{tabular}}} & recall@50                &  0.0716 (0.001)    & 0.0849 (0.002)&  0.0528 (0.002)        &  0.0827 (0.003) & 0.1086 (0.001) &  0.1034 (0.004) & 0.1123 (0.001)&     \textbf{0.1337 (0.002)}  & 19.1\%   \\ \cline{2-11} 
                                                                                           & \multicolumn{1}{c||}{MAP} &  0.0625 (0.002)   & 0.0654 (0.001) & 0.0517 (0.002)         &  0.0658 (0.001) & 0.0815 (0.004) & 0.0746 (0.005)  &0.0821 (0.003)&     \textbf{0.0934 (0.003)}     &13.8\% \\ \hline

\multirow{2}{*}{\textit{\begin{tabular}[c]{@{}c@{}}Amazon \\Movies and TV\end{tabular}}} & recall@50                &  0.0643 (0.001)    &0.0604 (0.002)&  0.0496 (0.002)      &  0.0711 (0.002)& 0.0935 (0.001)&  0.0901 (0.001)& 0.1001 (0.003)&     \textbf{0.1196 (0.002)}  & 19.5\%   \\ \cline{2-11} 
                                                                                           & \multicolumn{1}{c||}{MAP} &  0.0543 (0.001)   &  0.0551 (0.001)& 0.0493 (0.002)       &  0.0659 (0.004)& 0.0771 (0.002)& 0.0746 (0.004)  & 0.0785 (0.001)&     \textbf{0.0895 (0.002)}     &14.0\% \\ \hline
\end{tabular}
\end{table*}
\begin{figure*}[t]
\centering
\includegraphics[height=0.15\textheight,width=0.3\textwidth]{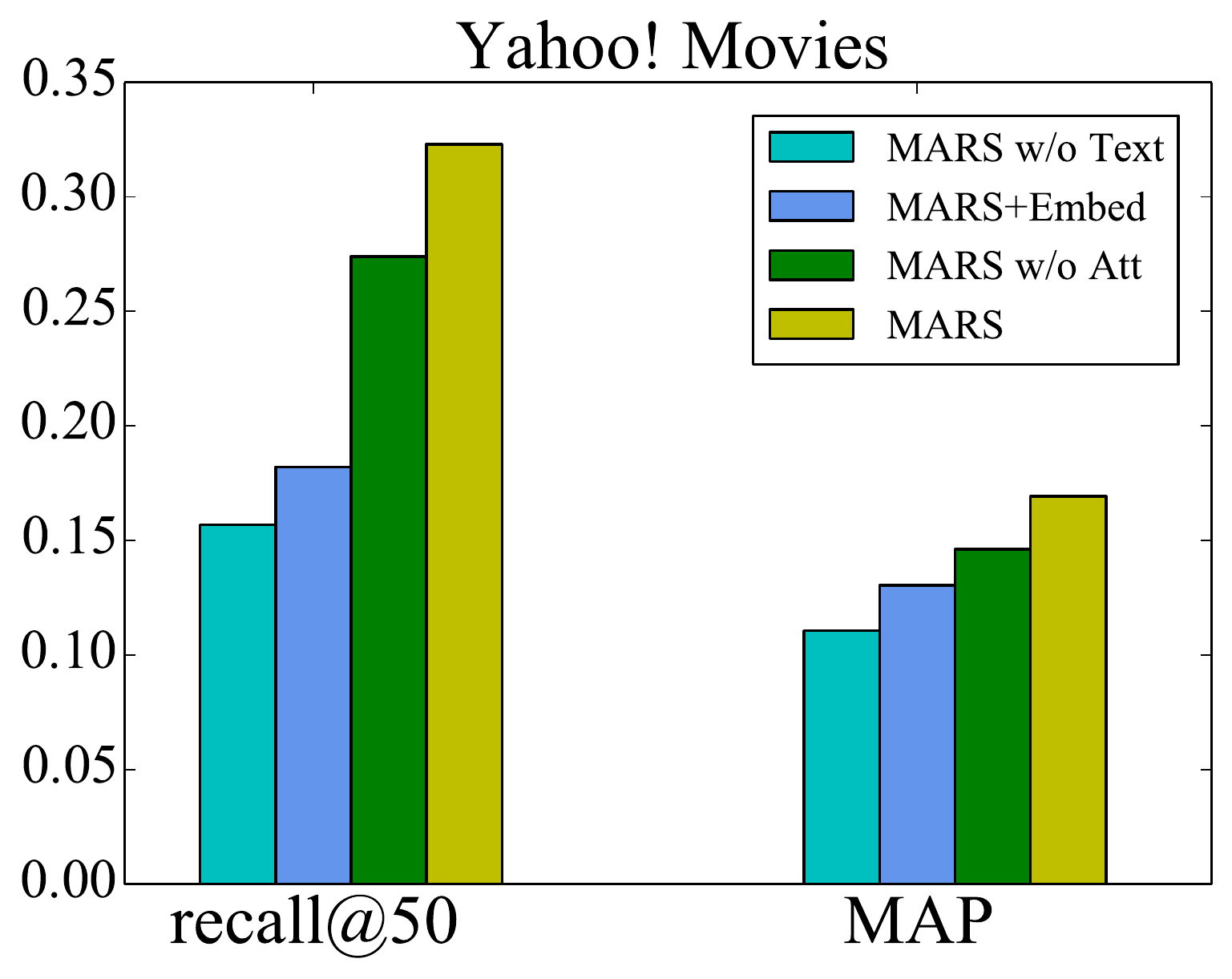}
\includegraphics[height=0.15\textheight,width=0.3\textwidth]{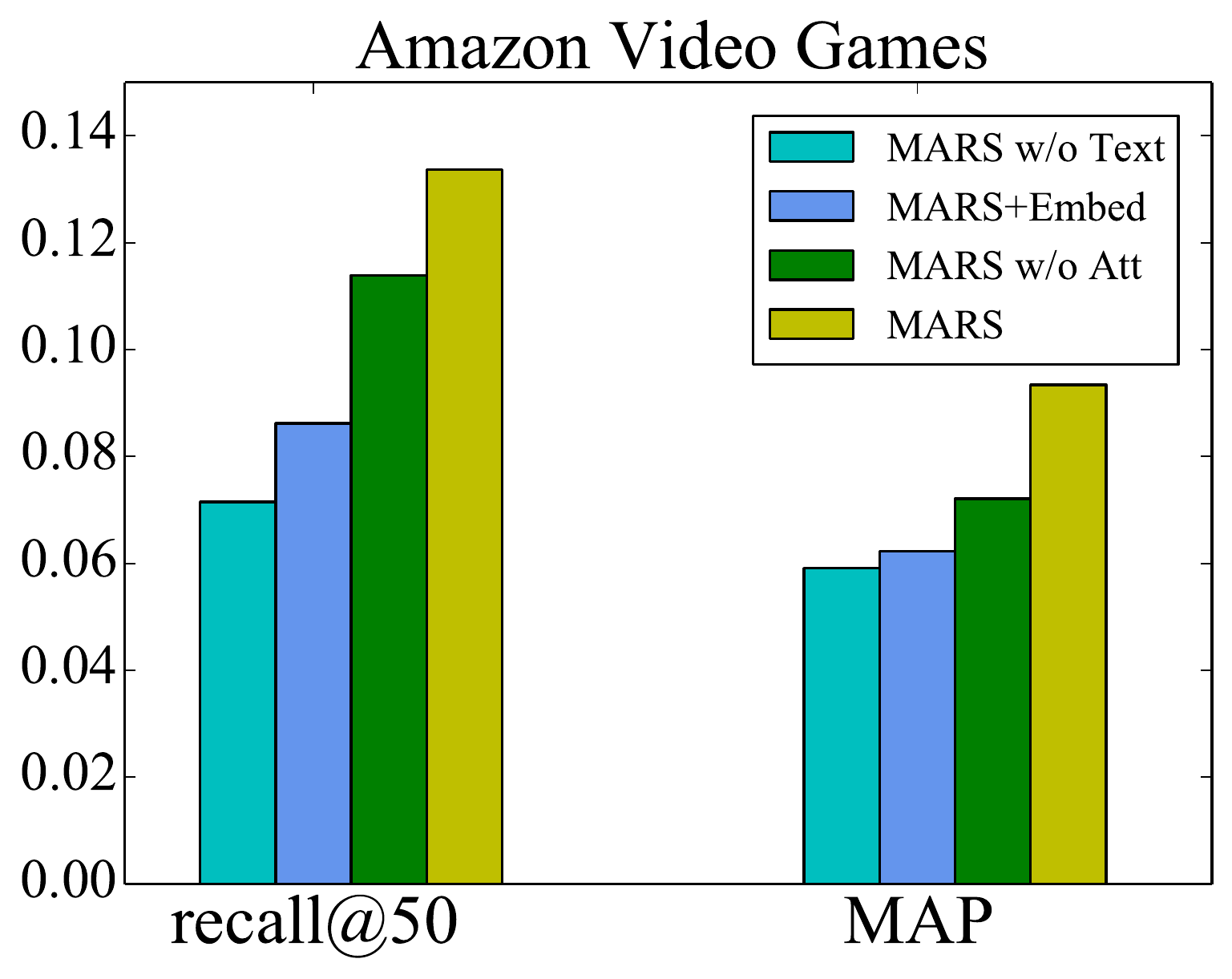}
\includegraphics[height=0.15\textheight,width=0.3\textwidth]{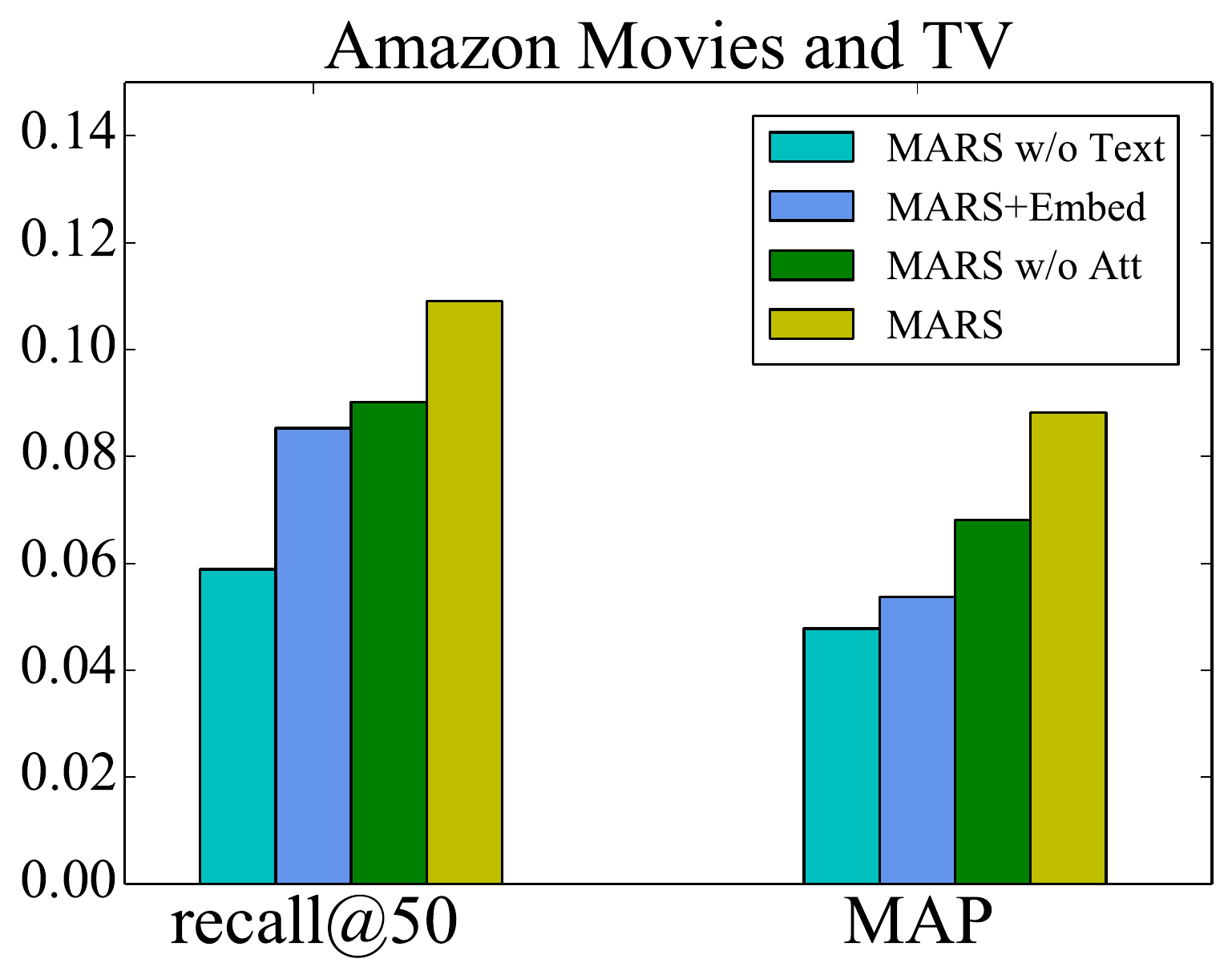}
\vspace{-1em}
\caption{Performance comparison with variants of MARS on the three datasets in terms of recall@$50$ and MAP. }
\label{invariants}
\centering
\end{figure*}

Table \ref{baseline} illustrates the experimental results for MARS and seven state-of-the-art baselines on all three datasets. Overall, MARS improves the best baseline by \textbf{20.8}\% and \textbf{13.0}\% in terms of recall@50 and MAP, respectively, averaging on all three datasets. 

These experiments reveal a number of interesting points:
\begin{itemize}
\item Regardless of the data sets and the evaluation metrics, our MARS always
achieves the best performance. This shows that by
leveraging the power of \textit{adaptive user representations},
MARS can better model users' diverse interests, resulting in better recommendations.
\item Besides CMF, models considering the texutal content generally give better results than models ignoring the textual content. It validates the usefulness of the textual content.
\item Deep models perform better than "shallow" ones in our experiments. This observation indicates the advantages of deep features extracted by various deep models.
\item Among all baseline models, Wide \& Deep shows itself as a strong performer and defeats all other baselines in all three datasets in terms of recall@50 and MAP. It calls us to combine a deep model with a "shallow" one for improvements.
\item Because of the sparsity differences, we also observe that the performances of all models degrade as the dataset become sparser.
\end{itemize}
\section{Model Analysis}
In this section, we conduct experiments to
answer the following questions:
\begin{itemize}[leftmargin=0cm]
\item[] \textbf{Q1}: How much does MARS benefit from taking the textual content associated with items into consideration?
\item[] \textbf{Q2}: How much do the incorporated CNN models help MARS?
\item[] \textbf{Q3}: Do the proposed \textit{adaptive user representations} and attention mechanism assist MARS in achieving better performances?
\end{itemize}

In order to answer questions above, we include three variants of MARS: MARS w/o Text, MARS+Embed and MARS w/o Att as the following:
\begin{itemize}
\item \textbf{MARS w/o Text}: To answer \textbf{Q1}, instead of learning item embeddings from the textual content, we ignore the content information and randomly initialize item embeddings based on a Gaussian distribution. After the initialization, embeddings of items are optimized during the training.
\item \textbf{MARS+Embed}: In order to answer \textbf{Q2}, in this variant, we replace the CNN models by simply averaging on embeddings of words contained each document associated with every item.
\item \textbf{MARS w/o Att}: For \textbf{Q3}, we include a variant of MARS without the proposed attention mechanism. To do so, we fix all values of $\boldsymbol\alpha$ as one. In this way, for a user $i$, the attention vector is canceled and representation of user $i$ becomes fixed and is not adaptive to relevant items liked by user $i$.
\end{itemize}
As shown in Figure \ref{invariants}, MARS w/o Text performs the worst, regardless of datasets and evaluation metrics. Compared with MARS w/o Text, MARS+Embed gains improvements in all three datasets. It further validates the advantages of incorporating the textual content for RS. However, averaging on embeddings of words associated with items is not an ideal method to make use of information existing in the textual content. Powered by deep features extracted by the CNN models from the textual content, MARS w/o Att achieves additional improvements over MARS+Embed. In terms of MAP, MARS w/o Att improves MARS+Embed by $12.0\%$, $15.73\%$ and $26.82\%$ on the dataset of \textit{Yahoo! Movies}, \textit{Amazon Video Games} and \textit{Amazon Movies and TV}, respectively. However, MARS defeats MARS w/o Att in terms of recall@50 and MAP on all three datasets. This demonstrates that, with the proposed attention mechanism and the incorporated CNN models, MARS is able to learn an effective \textit{adaptive user representation} and leverage the rich information existing in the textual content.
\section{A case study on the interpretability of MARS}
\begin{table*}[t]
\centering
\caption{A case study on the interpretability of MARS. The second column displays the top 2 movies recommended by MARS to \textit{User I} and \textit{User II}. The third column includes the top three movies with highest attention values. The actual attention value is shown in parentheses.}
\vspace{-1em}
\label{table::case_study}
\begin{tabular}{|M{1cm}||M{5cm}|l|}
\hline
        & \textbf{Recommended Movies}                                           & \multicolumn{1}{c|}{\textbf{Because you watched} }                                                                                                      \\ \hline
\textit{User I}  & \textit{Sleepless in Seattle}                          & \begin{tabular}[]{@{}l@{}}1. \textit{Where the Heart Is} (0.03)\\  2. \textit{When Harry Met Sally...} (0.02)\\  3. \textit{Ronnie \& Julie} (0.02)\end{tabular}                                                            \\ \cline{2-3} 
		& \textit{Enemy at the Gates}                          & \begin{tabular}[l]{@{}l@{}}1. \textit{X2: X-Men United} (0.02)\\  2. \textit{Top Gun} (0.015)\\  3. \textit{The Matrix Reloaded} (0.014)\end{tabular}                                                            \\ \cline{2-3}
        
\hline
\textit{User II}  & \textit{The Lord of the Rings: The Two Towers}                         & \begin{tabular}[]{@{}l@{}}1. \textit{Pirates of the Caribbean: The Curse of the Black Pearl} (0.03)\\  2. \textit{Harry Potter And the Chamber of Secrets}  (0.02)\\  3. \textit{Harry Potter and the Sorcerer's Stone} (0.02)\end{tabular}                                                            \\ \cline{2-3} 
		& \textit{Kill Bill Vol. 1}                        & \begin{tabular}[l]{@{}l@{}}1. \textit{Donnie Brasco} (0.03)\\  2. \textit{The Italian Job} (0.021)\\  3. \textit{S.W.A.T.} (0.02)\end{tabular}                                                            \\ \cline{2-3}   
\hline
\end{tabular}
\end{table*}
To gain a better insight into the interpretation ability of MARS, we conduct a qualitative experiment in this section. We run MARS on the dataset of \textit{Yahoo! Movies} and examine two example users. For each of them, we show two recommended movies in the second column of Table \ref{table::case_study}. And the top three movies with the highest attention values shown in the third column explain why corresponding movies in the second column are recommended by MARS. For example, MARS recommends \textit{Sleepless in Seattle} to \textit{User I} because movies, such as \textit{Where the Heart Is} and \textit{When Harry met Sally...}, are locally activated by high attention values. Besides \textit{romance} movies, MARS also discovers that \textit{User I} is interested in \textit{action} movies because of his or her past interactions with \textit{X2: X-Men United} and \textit{Top Gun}. As a result, MARS recommends \textit{Enemy at the Gates}. For \textit{User II}, MARS recommends two movies of different genres: \textit{The Lord of the Rings: The Two Towers} and \textit{Kill Bill Vol. 1 (2003)}. It is because movies watched by \textit{User II}, such as \textit{Pirates of the Caribbean: The Curse of the Black Pearl} and \textit{Donnie Brasco (1997)}, indicate the user is a fan of \textit{fantasy} and \textit{action} movies. Overall, this case study shows that MARS is not only able to capture users' diverse interests but also has a great interpretability to explain its recommendation results.
\section{Related Work}
\label{sec::related}
Our work is closely related to two areas: deep learning based RS and attentional mechanisms employed for RS. We will first give a brief review on deep learning based RS. Then, we covers works utilizing attentional mechanisms for RS.
\subsection{Deep Recommender Systems}
Several studies \cite{zheng2016neural,wang2017irgan,lu2018learning} recently propose deep learning based models for the recommendation tasks. One pioneer work \cite{salakhutdinov2007restricted} in this area uses a Restricted Boltzmann Machines (RBM) \cite{hinton2010practical} based method to model users using their rating preferences. Followed by this trend, \cite{wu2016collaborative} utilize denoising Auto-encoders to learn latent vectors of users and items from the rating matrix. \cite{he2017neural} and \cite{NFM} leverage Multilayer Perceptron (MLP) to learn from user-item interactions. In \cite{zheng2016neural}, a CF-based Neural Autoregressive Distribution Estimator (CF-NADE) model is proposed for collaborative filtering tasks. However, all works above differ from MARS because they ignore the rich content associated with items. 

Some studies also propose to utilize deep learning techniques to build a content-based recommender system. In \cite{van2013deep,wang2014improving}, authors introduce Convolutional Neural Networks (CNN) \cite{lecun1990handwritten} and Deep Belief Network (DBN) \cite{hinton2006fast} to learn users' preferences from music data. \cite{wang2015collaborative,wang2016collaborative,bansal2016ask} propose a deep recommendation model learning from the textual content associated with items. \cite{covington2016deep,wang2017your} propose deep recommender systems for video and point-of-interest recommendation. \cite{zhang2016collaborative, zhang2017joint,zheng2017joint} investigate how to leverage the multi-view information to improve the quality of recommender systems. In \cite{li2017neural}, they propose a model which is able to simultaneously
predict ratings and generate abstractive tips. For more works on the deep learning based RS, readers can refer to a survey paper \cite{zhang2017deep}.

\subsection{Attentional Mechanisms for RS}
Recently, attentional mechanisms attract considerable interests of researchers, owing to their ability of modeling users' attention. A number of works \cite{vinh2018attention} employing attentional mechanisms to build RS have been proposed. \cite{chen2018neural} introduces an attentional mechanism utilizing reviews. In \cite{xiao2017attentional}, authors propose a neural attention network to model the importance of each feature interaction
from data. \cite{loyola2017modeling} proposed an attentional mechanism based model to learn dynamics of user interests from a sequence of items. To build an interpretable recommendation model, \cite{seo2017interpretable} proposes to utilize an attentional method to extract text form reviews. For the purpose of capturing editors' dynamic criteria for selecting news articles, \cite{wang2017dynamic} proposes a dynamic attentional method.

To the best of our knowledge, two methods close to ours are presented in \cite{chen2017entive} and \cite{zhou2017deep}. Although the two works proposed to place attentional mechanisms on items, none of them are end-to-end models. It means that they use either hand-crafted or pre-trained features. As a result, item features are not optimized for the task of recommendation. It leads to an inaccurate attentional mechanism. In contrast, MARS learns item features and users' attention in an end-to-end fashion. Therefore, compared with them, MARS achieves better item features and more accurate attention of users. 

In terms of interpretability, a deep recommender system with interpretability is presented in \cite{seo2017interpretable}. However, they interpret recommendations based on reviews written by users. In contrast, we interpret recommendations based on items purchased/liked by users. In fact, interpreting recommendations based on purchased items is more popular in real life (as \textit{Amazon} or \textit{Netflix} does).

Although all neural network based approaches above are deep content-based recommender systems, they differ from MARS because all of works above are unable to learn an \textit{adaptive user representation} in an end-to-end fashion.
\section{Conclusions}
\label{sec::con}
Although deep learning based RS have shown promising results in a variety of recommendation tasks, previous methods often focus on utilizing deep models for modeling item representations and learning a fixed user representation. However, a fixed user representation restrains models from modeling diverse interests of users. Furthermore, while interpretability is demanded in many recommendation scenarios, most of existing deep learning based RS are unable to interpret their recommendation results.

In this paper, to tackle the problems and challenges above, we present a Memory Attention-aware Recommender System (MARS) model. With a proposed memory component and an item-level attention mechanism, instead of modeling fixed deep user representations, MARS learns a deep \textit{adaptive user representation}. For an item $j$ and a set of items liked by user $i$, a deep \textit{adaptive user representation} can dynamically adapt to those items in the set which are relevant item $j$. Owing to its adaptability, a deep \textit{adaptive user representation} can overcome the difficulty of modeling users' diverse interests. Moreover, with the help of the proposed attention mechanism, MARS is able to interpret its recommendation results based on purchased items of users. 

In the experiments, we demonstrate that MARS achieves superior performances by comparing with seven state-of-the-art methods on three real-world datasets. Also, we demonstrate that MARS can not only overcome the difficulty of modeling diverse interests of users but also has a great interpretability.
\bibliographystyle{ACM-Reference-Format}
\bibliography{bib1.bib} 

\end{document}